\documentclass[%
 reprint,
 amsmath,amssymb,
 aps,
]{revtex4-2}
\usepackage[version=3]{mhchem}
\usepackage{graphicx}
\usepackage{dcolumn}
\usepackage{bm}

\begin{document}

\preprint{APS/123-QED}

\title{Altermagnetic ferroelectric LiFe$_2$F$_6$ and spin-triplet excitonic insulator phase}

\author{Peng-Jie Guo$^{1}$}
\email{guopengjie@ruc.edu.cn}
\author{Yuhao Gu$^{2}$}
\email{guyuhao0709@gmail.com}
\author{Ze-Feng Gao$^{1}$}
\author{Zhong-Yi Lu$^{1}$}
\email{zlu@ruc.edu.cn}
\affiliation{1. Department of Physics, Renmin University of China, 100872, Beijing, China}

\affiliation{2. Department of Physics, University of Science and Technology Beijing, 100083, Beijing, China}

\date{\today}

\begin{abstract}
Altermagnetism is a new magnetic phase with $k$-dependent spin polarization and may exist in an insulatinging state with a high Néel temperature. This provides a new opportunity to obtain both spin and electric polarization in one material. Here, based on symmetry analysis and the first-principles electronic structures calculations, we predict that the \ce{LiFe2F6} is a $d$-wave altermagnetic and charge-ordering-mediated ferroelectric material. Moreover, the \ce{LiFe2F6} transforms into a ferrimagnetic and ferroelectric phase with strong magnetoelectric coupling under biaxial compressive strain. Interestingly, the spins of the valence band and the conduction band are opposite in ferrimagnetic \ce{LiFe2F6}, which facilitates a simutaneously spin-triplet excitonic insulator phase. More importantly, the spin triplet excitons with spin 1 and -1 can be switched by electric fields in ferrimagnetic \ce{LiFe2F6} due to strong magnetoelectric coupling. Due to the abundance of novel physical properties, \ce{LiFe2F6} will certainly attract a wide range of theoretical and experimental interest.
\end{abstract}

\maketitle


{\it Introduction.} Multiferroics with the coexistence of ferroelectric order and magnetic order are one of the cores of condensed matter physics and may be used in the next generation electronic devices~\cite{nature-mul, NM-mul, dong-review,NRM-mul, NM-mulferro}. Usually, ferromagnetism takes place in metals, while ferroelectric materials  are insulators. Thus, multiferroic materials with coexisting magnetization and polarization are very rare~\cite{prl-mulferro, nature-EuTiO}. Most of the multiferroic materials have coexisting antiferromagnetic and ferroelectric phases. In addition, there are a few ferrimagnetic materials with coexisting magnetization and polarization~\cite{prl-rerri, PRM-LiFeF}. 

Very recently, altermagnetism as a new magnetic phase has been theoretically proposed and experimentally verified to be distinct from ferromagnetism and conventional antiferromagnetism~\cite{li-prx, liber-prx, Bai-PRL, Ka-PRL, MnTe-PRL, Feng-NE}. Due to the absence of both spin symmetry $\{T\|IT\}$ and $\{C_{2}^{\bot}\|t\}$, the altermagnetic materials have $k$-dependent spin polarization, which can lead to $d$-wave, $g$-wave, and $i$-wave magnetic phase depending on the spin group symmetry~\cite{li-prx}. Here, the symmetry operations at the left and right of the double vertical bar act only on the spin space and lattice space, respectively; the notation $C_{2}^{\bot}$ represents the 180 degrees rotation perpendicular to the spin direction; the notations $I$, $T$ and $t$ denote space-inversion, time-reversal, and fractional translation operations, respectively. The $k$-dependent spin polarization in altermagnetic materials can result in many novel physical effects, such as the unique spin current~\cite{Liber-PRL, Bai-PRL, Ka-PRL}, the giant magnetoresistance, the tunneling magnetoresistance~\cite{liber-prxm} and nontrivial superconductivity~\cite{topological-prb}. Moreover, like ferromagnetic materials, altermagnetic materials also break time-reversal symmetry. Therefore, the time-reversal symmetry-breaking macroscopic phenomena, such as quantum anomalous Hall~\cite{Guo2023}, anomalous Hall~\cite{CAHE-2020, MnTe-PRL, Feng-NE, Liber-NRM, hou-prb, gao2023}, anomalous Kerr effects~\cite{Zhou-PRB} and so on, can be also realized in altermagnetic materials. But different from ferromagnetism, altermagnetism may exist in an insulating state with high a Néel temperature. Accordingly, altermagnetism and ferroelectricity can be realized in a single material, which will open up new opportunities to achieve spin polarization and ferroelectric polarization simultaneously in a single material. However, so far, only one candidate material $CaMnO_3$ has $d$-wave altermagnetism and ferroelectricity~\cite{liber-prx}. Thus, predicting more materials with coexisting altermagnetism and ferroelectricity is very urgent for the study of their novel physical properties. 

On the other hand, excitonic insulators need to satisfy that the excitonic bonding energy is larger than the single-particle energy gap and the excitons have spontaneous Bose condensation~\cite{RMP-1, RMP-2}. Although much progress has been made in this area~\cite{sci-ex, Nc-ex, Nc-exci}, compelling experimental evidence is still lacking. This is because excitons are electrically neutral, resulting in no suitable experimental means to detect the exciton condensation for nonmagnetic excitonic insulators. Recently proposed spin-triplet excitonic insulators have spin signal, which can be detected by spin transport experiments. The spin-triplet excitonic state requires that the spins of the conduction band and the valence band of ferromagnetic insulator are opposite and with weak spin-orbit coupling (SOC)~\cite{jiang-prl}. Moreover, due to the spin selection rule, this type of ferromagnetic insulators favor excitonic insulators~\cite{jiang-prl}. Unfortunately, so far, this type of ferromagnetic insulators have not yet been discovered experimentally. The altermagnetism and ferrimagnetism with opposite spins arrangement may provide new directions for finding magnetic materials with opposite spin of the conduction band and the valence band.  

In this work, based on symmetry analysis and the first-principles electronic structures calculations, we predict that the \ce{LiFe2F6} is a multiferroic material with $d$-wave altermagnetism and ferroelectricity. Then, we investigate the physical properties of \ce{LiFe2F6} under biaxial strain. We find that the \ce{LiFe2F6} changes from $d$-wave altermagnetic ferroelectric phase to ferrimagnetic ferroelectric phase. Interestingly, the spins of the valence band and the conduction band are opposite in ferrimagnetic \ce{LiFe2F6}, which facilitates a spin-triplet excitonic insulator phase. More importantly, the spins of the bottom conduction band and the top valence band can be switched by electric field due to strong magnetoelectric coupling.

{\it Method.} Our electronic structure calculations employed the Vienna ab initio simulation package (VASP) code~\cite{kresse1996} with the projector augmented wave (PAW) method~\cite{Joubert1999}. The Perdew-Burke-Ernzerhof for solids (PBEsol)~\cite{perdew2008} exchange-correlation functional and the GGA plus on-site repulsion $U$ method (GGA+$U$) in the formulation of Liechtenstein \textit{et al.}~\cite{liechtenstein1995} were used in our calculations. Here, the effective on-site exchange interaction $J$ was fixed to U/5. Moreover, the screened hybrid functional~\cite{Jcp-1, Jcp-2} introduced by Heyd, Scuseria, and Ernzerhof (HSE) with the HSE06 version~\cite{Jcp-3} were also used to calculate the electronic structure. We used the standard Berry phase method to estimate the ferroelectric polarization \textbf{P}~\cite{king1993, resta1994}. We adopted the nudged elastic band (NEB)~\cite{NEB} method to simulate the flipping of \textbf{P} and estimate the energy barrier in this process. The kinetic energy cutoff was set to be 650 $eV$ for the expanding the wave functions into a plane-wave basis and the energy convergence criterion was $10^{-7}$ $eV$. The crystal structures were fully relaxed until the force on each atom was less than 0.001 $eV$/\AA. The $\Gamma$-centered \textbf{k}-mesh was set as $16\times16\times8$. In the calculations of strained \ce{LiFe2F6}, the in-plane lattice constants were fixed while the length of the $c$ axis and the atomic positions were optimized.

\begin{figure}[t]
\centering
\includegraphics[width=0.45\textwidth]{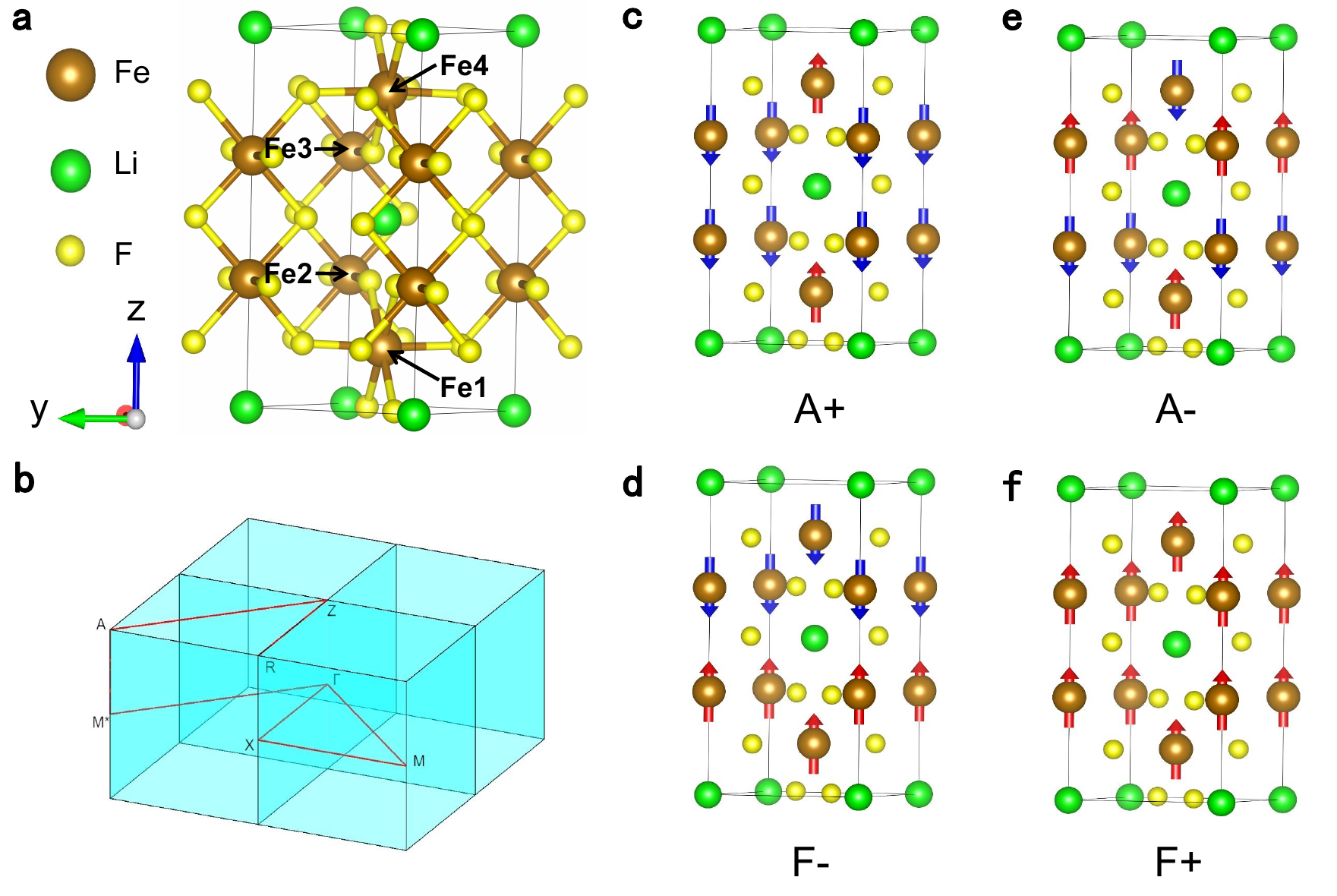}
\caption{Crystal structure, Brillouin zone (BZ) and four most relevant collinear magnetic structures of \ce{LiFe2F6}. 
{\bf{a}}, Crystal structure of Li2FeF6 with high symmetry $P4_2-mnm~ (136)$. 
{\bf{b}}, the corresponding BZ with high-symmetry point and line. 
{\bf{c}--\bf{f}}, The A+, A-, F- and F+ collinear magnetic structure of \ce{LiFe2F6}, respectively. 
The red and blue arrows represent spin-up and spin-down magnetic momentums, respectively.}
\label{fig:1}
\end{figure}
\begin{figure}[t]
\centering
\includegraphics[width=0.45\textwidth]{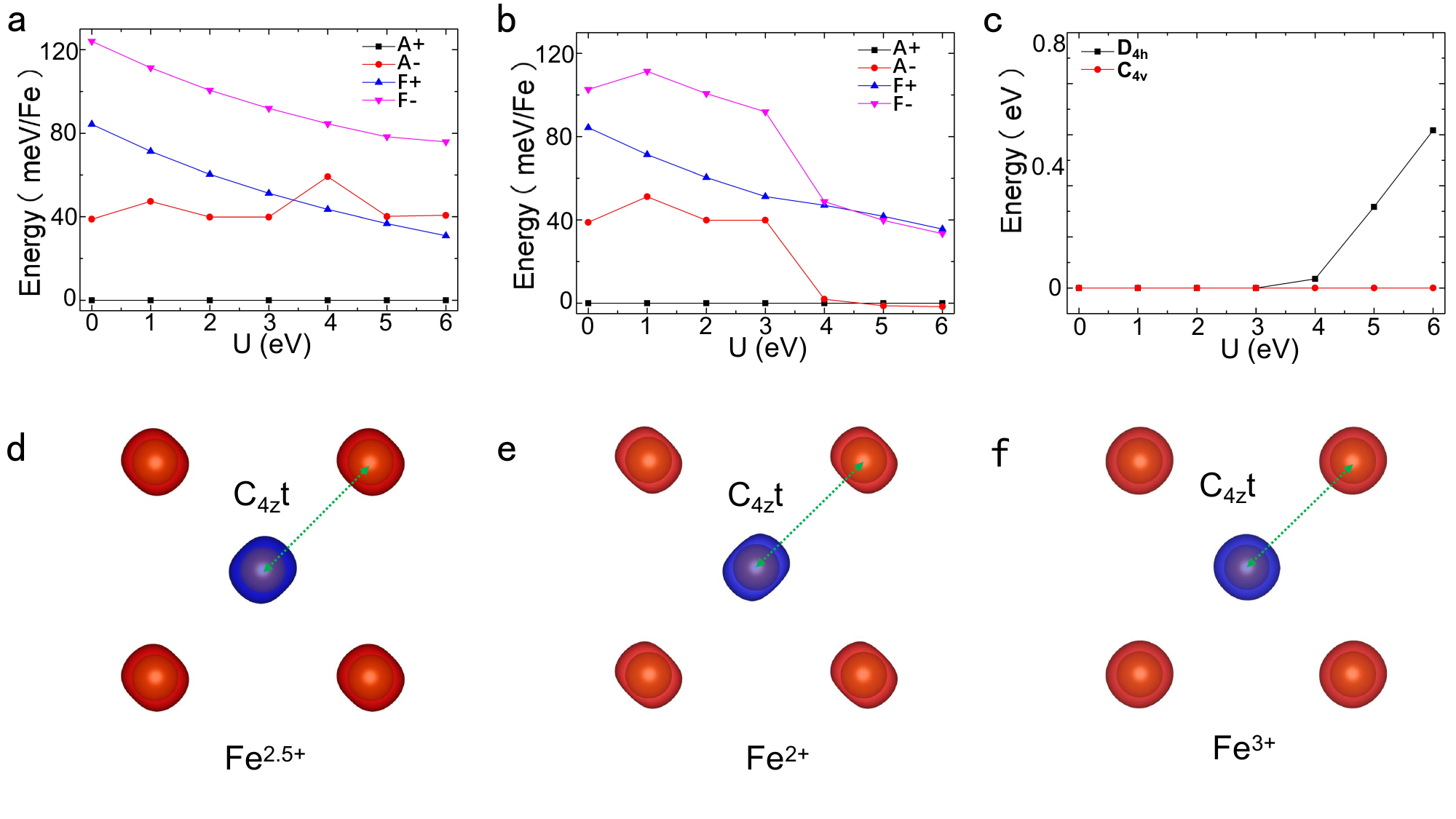}
\caption{Relative energy of different magnetic states with the variation of correlation interaction $U$ and polarization charge density for \ce{LiFe2F6}. 
{\bf{a}}, The relative energy of different magnetic states with the variation of correlation interaction $U$ for high-symmetry $P4_2-mnm~(136)$ phases.
{\bf{b}}, The relative energy of different magnetic states with the variation of correlation interaction $U$ for low-symmetry $P4_2nm~(102)$ phases.
{\bf{c}}, The relative energy of different crystal phase with the variation of correlation interaction $U$. 
{\bf{d}--\bf{f}}, The polarization charge density of $Fe^{2.5+}, Fe^{2+}$ and $Fe^{3+}$, respectively. 
The red and blue represent spin-up and spin-down charge density, respectively. 
The polarization charge densities are calculated under correlation interaction $U = 4~eV$ and exchange interaction $J=0.8~eV$. 
The $t$ represents a fractional translation with (1/2, 1/2, 1/2). 
}
\label{fig:2}
\end{figure}

{\it Results and discussion.} \ce{LiFe2F6} has a tetragonal crystal structure (Fig.~\ref{fig:1}{\bf{a}}). At high temperature, \ce{LiFe2F6} has $P4_2/mnm~(136)$ space group symmetry. The corresponding Brillouin zone (BZ) is shown in Fig.~\ref{fig:1}{\bf{b}} and the high-symmetry lines and points are labeled. Due to the nonsymmorphic space symmetry, there are four $Fe$ atoms in the primitive cell of \ce{LiFe2F6}. From Fig.~\ref{fig:1}{\bf{a}}, the four $Fe$ atoms can be divided into two classes according to different orientations of $Fe-F$ octahedrons, example for $Fe1$ and $Fe2$. According to the four $Fe$ atoms in primitive cell, there are the most relevant collinear magnetic structures including three collinear antiferromagnetic states A+, A-, F- and ferromagnetic state $F+$ (Fig.~\ref{fig:1}{\bf{c}--\bf{f}}). Since the angles of $Fe1-F-Fe2$ and $Fe2-F-Fe3$ are respectively 133 and 95 degrees, the superexchange interactions may lead to $Fe2$ $(Fe3)$ and $Fe1 (Fe4)$ with opposite spin arrangement and $Fe1 (F2)$ and $Fe4 (Fe3)$ with the same spin arrangement, which corresponds to A+ antiferromagnetic state. To determine the magnetic ground state of the high-symmetry \ce{LiFe2F6}, we calculate relative energies of four different magnetic states with the variation of correlation interaction $U$. From Fig.~\ref{fig:2}{\bf{a}}, the A+ antiferromagnetic state is the most stable, which is consistent with our theoretical analysis.

For A+ antiferromagnetic state, the $Fe$ ions with the same spin arrangement are connected by $I$, thus the spin symmetry $\{C_{2}^{\bot}\|I\}$ is broken. Moreover, the spin symmetry $\{C_{2}^{\bot}\|t\}$ is also broken due to nonmagnetic $F$ anions. Considering that the $Fe$ ions with opposite spin arrangement can be connected by the spin symmetry $\{C_{2}^{\bot}\|C_{4z}t\}$, the A+ antiferromagnetic state is a $d$-wave altermagnetic state. In order to show the $d$-wave altermagnetic characteristics more intuitively, we calculate the polarization charge density of the high-symmetry \ce{LiFe2F6}. From Fig.~\ref{fig:2}{\bf{d}}, spin-up and spin-down $Fe$ ions have anisotropic polarization charge density deriving from the $Fe-F$ octahedrons of different orientations. Obviously, the $Fe$ ions with opposite spin polarization are not connected by the $I$ or $t$ transformation, but connected by the $C_{4z}t$ transformation. 

By analyzing the chemical formula of \ce{LiFe2F6}, all $Fe$ ions should be of 2.5 valence. Indeed, our calculations show that all $Fe$ ions are of 2.5 valence in the high-symmetry \ce{LiFe2F6}. On the other hand, Mössbauer experiment found that there are $Fe^{2+}$ and $Fe^{3+}$ in \ce{LiFe2F6}~\cite{jcsa-1} and X-ray diffraction experiment revealed a low-symmetry $P4_2nm~(102)$ phase above room temperature~\cite{jssc-2}. Thus, \ce{LiFe2F6} may exist a ferroelectric phase transition induced by charge order, which has been demonstrated by Dong's theoretical calculations~\cite{PRM-LiFeF}. Since there is only a slight difference between the high-symmetry and the low-symmetry structures, the A+ altermagnetic state may be still most stable in the low-symmetry \ce{LiFe2F6}, which has been confirmed by previous neutron scattering experiment~\cite{nds-LiFeF}. Thus, \ce{LiFe2F6} is a multiferroic material with $d$-wave altermagnetism and ferroelectricity. In order to show the altermagnetic and charge ordering characteristics  more intuitively, we also calculate the polarization charge density of the low-symmetry \ce{LiFe2F6}, as shown in Fig.~\ref{fig:2}{\bf{e}} and {\bf{f}}. The large difference of polarization charge density of $Fe^{2+}$ and $Fe^{3+}$ reflects the charge order (Fig. ~\ref{fig:2}{\bf{e}} and {\bf{f}}). Comparing the polarization charge density of $Fe$ ions with different valence states, the anisotropy of polarization charge density of $Fe^{2+}$ is the strongest, while that of $Fe^{3+}$ is the weakest. Obviously, the $Fe^{2+}$ ($Fe^{3+}$) ions with opposite spin polarization are not connected by the $I$ or $t$ transformation, but connected by the $C_{4z}t$ transformation, which reflects $d$-wave altermagnetic characteristic.

In order to further investigate the properties of altermagnetism and ferroelectricity, we need to determine a suitable correlation interaction $U$. This suitable correlation interaction $U$ value can be determined by the existing experimental results and our calculation results. Then, we calculate relative energies of four different magnetic states with the variation of correlation interaction $U$ for the low-symmetry \ce{LiFe2F6}. Different from the high-symmetry \ce{LiFe2F6}, the most stable magnetic state changes from A+ altermagnetic state to A- ferrimagnetic state with increasing of correlation interaction $U$. Since the magnetic ground state is A+ altermagnetic state, the correlation interaction $U$ is less than 4.6 $eV$ (Fig.~\ref{fig:2}{\bf{b}}). On the other hand, when the correlation interaction $U$ is less than 3 $eV$, the charge order is not stable (Fig.~\ref{fig:2}{\bf{c}}). Thus, the correlation interaction $U$ is between 3 $eV$ and 4.6 $eV$. In the following calculation, we choose the correlation interaction $U$ to be equal to 4 $eV$.

\begin{figure}[t]
\centering
\includegraphics[width=0.45\textwidth]{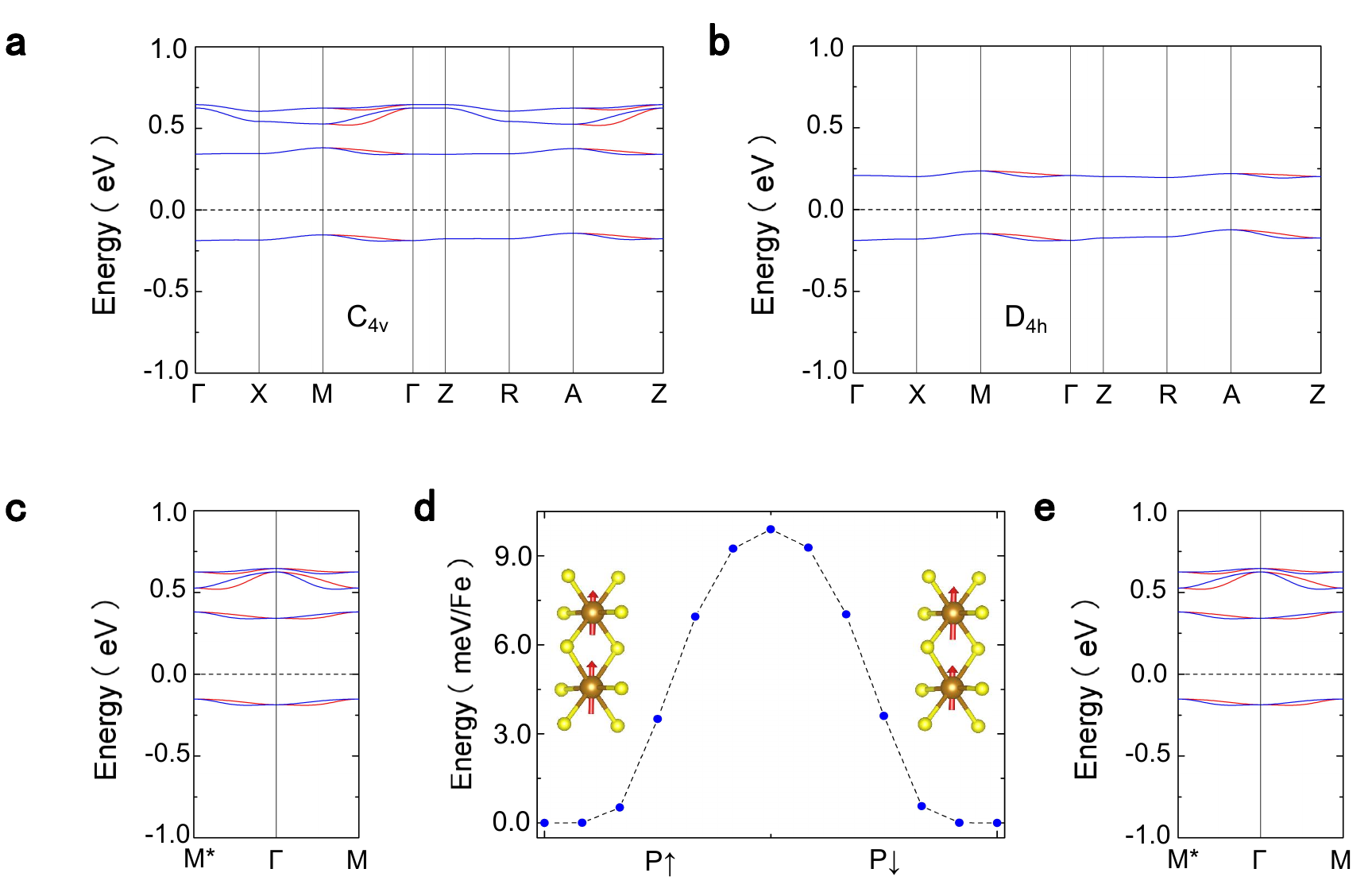}
\caption{
The electronic band structure without SOC and ferroelectric properties of \ce{LiFe2F6}. 
{\bf{a}}, The electronic band structures along the high-symmetry directions of low-symmetry \ce{LiFe2F6}, respectively. 
{\bf{b}}, The electronic band structures along the high-symmetry directions of high-symmetry \ce{LiFe2F6}, respectively. 
{\bf{c}} and {\bf{e}} are the electronic band structures with opposite ferroelectric polarization for low-symmetry \ce{LiFe2F6}, respectively. 
{\bf{d}}, The ferroelectric polarization simulated by the NEB method. 
Insets: Initial and final structures. 
The red arrows represent spin magnetic momentum. The long arrow is $Fe3+$ and the short arrow is $Fe2+$. 
The red and blue lines represent spin-up and spin-down bands. 
}
\label{fig:3}
\end{figure}
The electronic band structures were calculated for the low-symmetry and high-symmetry \ce{LiFe2F6}, which are shown in Fig.~\ref{fig:3}{\bf{a}} and {\bf{b}}. From Fig.~\ref{fig:3}{\bf{a}} and {\bf{b}}, both the high-symmetric and low-symmetric phases are semiconductors. Ferroelectric phase transition induced by charge order increases bandgap of \ce{LiFe2F6} from 311 $meV$ to 474 $meV$. Due to the absence of spin symmetry $\{C_{2}^{\bot}\|I\}$ and $\{C_{2}^{\bot}\|t\}$, altermagnetic \ce{LiFe2F6} has $k$-dependent spin splitting, example for the $\Gamma-M$ direction (Fig.~\ref{fig:3}{\bf{a}} and {\bf{b}}). In fact, due to the spin symmetry $\{C_{2}^{\bot}\|M_{x}t\}$ and $\{C_{2}^{\bot}\|M_{y}t\}$, spin-up and spin-down bands are degenerate in these four cyan faces (Fig.~\ref{fig:1}{\bf{b}}). Except for these four cyan faces, spin-up and spin-down bands are split at general $k$ points in the BZ. However, ferroelectric phase transition has only a small effect on spin splitting of the bands (Fig.~\ref{fig:3}{\bf{a}} and {\bf{b}}). Meanwhile, we also calculate the electronic band structure along the $M^*-\Gamma-M$ directions as shown in Fig.~\ref{fig:3}{\bf{c}}. From Fig.~\ref{fig:3}{\bf{c}}, the spin-up bands on the $M^*-\Gamma$ axis change into spin-down bands on the $\Gamma-M$ axis reflecting the characteristics of the $d$-wave altermagnetism. Considering the $d$-wave altermagnets favor unique spin current by electrical means, the low-symmetry \ce{LiFe2F6} may have spintronic, transistor and ferroelectric functionalities simultaneously. 

Different from traditional ferroelectrics, the ferroelectrics in \ce{LiFe2F6} origins from charge order, as $Fe^{2+}$ and $Fe^{3+}$ alternatively arrange. We used Berry phase method to estimate the ferroelectric P of \ce{LiFe2F6}, which gives 15.1 $\mu C/cm^2$ along the $z$ axis for the $d$-wave altermagnetic state. This is basically consistent with the intuitive charge order result, 12.3 $\mu C/cm^2$. Meanwhile, we also adopted NEB method to simulate ferroelectric polarization of \ce{LiFe2F6}. The energy barrier is only 9.9 $meV$ per $Fe$ atom (Fig.~\ref{fig:3}{\bf{d}}), which is substantially smaller compared to other ferroelectric materials. This seems natural as the main process of ferroelectric switching in \ce{LiFe2F6} is the charge transfer in $Fe^{2+}$-$Fe^{3+}$ pair, rather than the moving of the ions. On the other hand, the $Fe^{2+}$ and $Fe^{3+}$ ions in altermagnetic state have the same spin arrangement, the charge transfer in $Fe^{2+}$-$Fe^{3+}$ pair has no effect on the A+ altermagnetic state. Thus, the electronic band structures of polarized/antipolarized are the same, reflecting the weak magnetoelectric coupling in altermagnetic \ce{LiFe2F6} (Fig.~\ref{fig:3}{\bf{c}} and {\bf{e}}). However, in A- ferrimagnetic and F- altermagnetic state, the switch between positive and negative ferroelectric polarization can cause obvious change for the electronic structure as the spins in $Fe^{2+}$ and $Fe^{3+}$ pair are opposite. Note: The F- magnetic state in high-symmetry phase is a conventional collinear antiferromagnetic state, but the charge-order-induced ferroelectric phase transition will transform the F- from the conventional collinear antiferromagnetic state to an altermagnetic state, which is the reason of its strong magnetoelectric coupling.


\begin{figure}[t]
\centering
\includegraphics[width=0.45\textwidth]{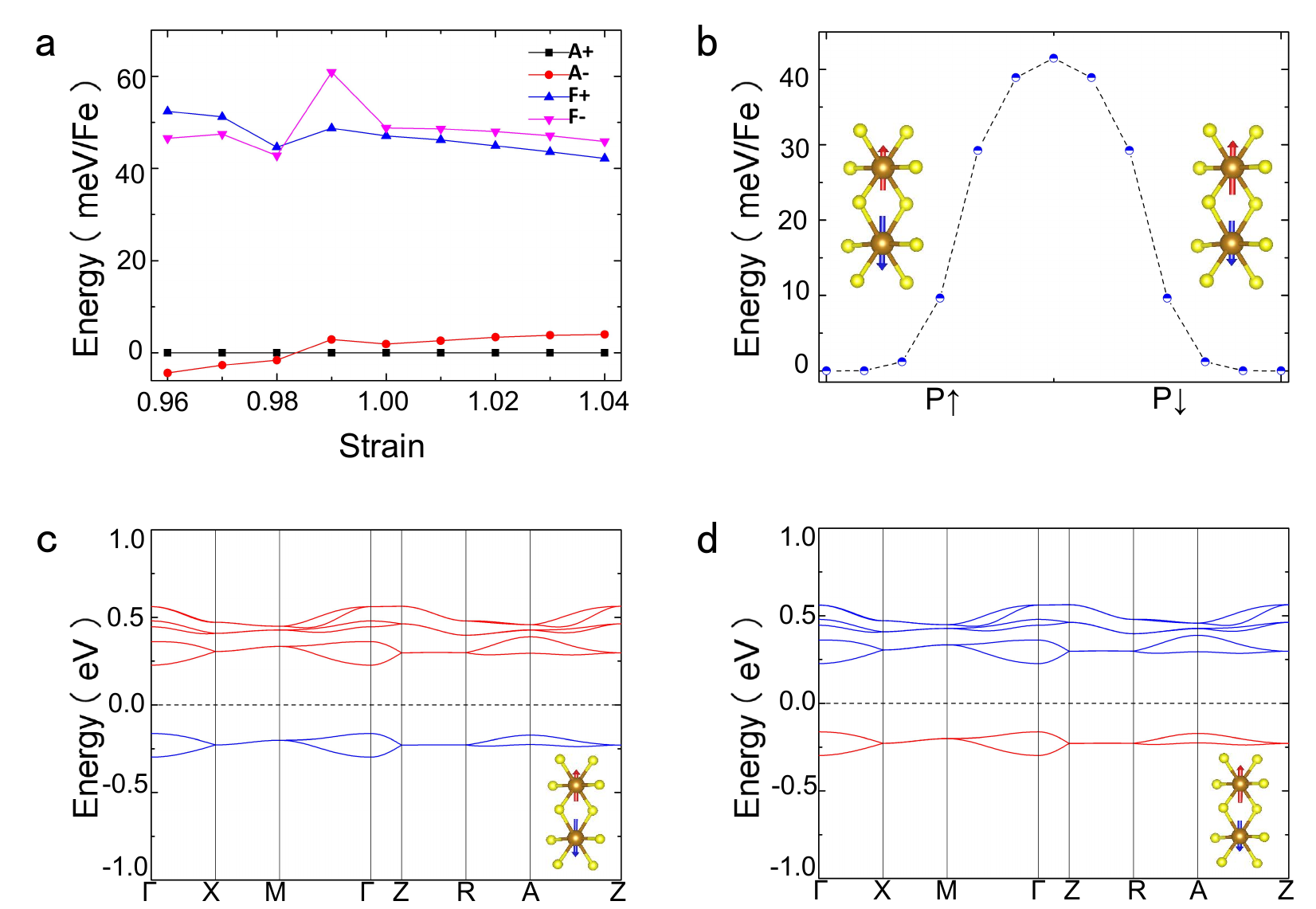}
\caption{
The results of low-symmetry \ce{LiFe2F6} under biaxial strain. 
{\bf{a}}, The relative energy of different magnetic states as a function of biaxial strain. 
{\bf{b}}, The NEB method is used to simulate ferroelectric polarization of ferrimagnetic \ce{LiFe2F6}. 
Insets: initial and final structures. 
{\bf{c}} and {\bf{d}} are the electronic band structures with opposite ferroelectric polarization of ferrimagnetic \ce{LiFe2F6}, respectively. 
The red and blue arrows represent spin-up and spin-down magnetic momentum, respectively. 
The long arrow is $Fe3+$ and the short arrow is $Fe2+$. 
The red and blue lines represent spin-up and spin-down bands.
}
\label{fig:4}
\end{figure}
{\it Possible spin-triplet excitonic insulator phase.} Interestingly, the low-symmetry \ce{LiFe2F6} can change from the A+ altermagnetic state to A- ferrimagnetic state under compressive biaxial strain~\cite{PRM-LiFeF}. We also calculate relative energies of four different magnetic states as functions of biaxial strain. Indeed, the altermagnetic state change to ferrimagnetic state under compressive biaxial strain (Fig.~\ref{fig:4}{\bf{a}}). Likewise, we also used Berry phase method to estimate the ferroelectric P which is 13.4 $\mu C/cm^2$ along the $z$ axis for the ferrimagnetic state, which is basically consistent with the previous calculation. The energy barrier is 41.5 $meV$ per Fe atom for the ferrimagnetic state (Fig.~\ref{fig:4}{\bf{b}}). For the positive ferroelectric polarization, the $Fe^{2+}$ and $Fe^{3+}$ ions have opposite spin arrangement, which makes the magnetic moment of a primitive cell to be -2 $\mu_B$. When the ferroelectric polarization is reversed, the magnetic moment of a primitive cell changes to 2 $\mu_B$ (Fig.~\ref{fig:4}{\bf{b}} Insets: final structure). Thus, the ferrimagnetic \ce{LiFe2F6} has very strong magnetoelectric coupling. Furthermore, the switch between positive and negative ferroelectric polarization may cause huge change in the electronic band structure. 

Then, we calculate the electronic band structures of the positive and negative ferroelectric polarization for ferrimagnetic \ce{LiFe2F6}. Comparing the positive and negative ferroelectric polarization, their electronic band structures are the same but the spin of bands is reversed (Fig.~\ref{fig:4}{\bf{c}} and {\bf{d}}), which corresponds to the reversal of the spin magnetic moment before and after the ferroelectric polarization reversal. Interestingly, the spins of the valence band and the conduction band are opposite (Fig.~\ref{fig:4}{\bf{c}} and {\bf{d}}). Since $Li$, $Fe$, and $F$ are all light elements, the \ce{LiFe2F6} has weak SOC. In the ferrimagnetic \ce{LiFe2F6} with weak SOC, the transition of electrons obeys the spin selection rule. Thus, electrons transition from the bottom valence band to the top conduction band need spin flip, the electron-hole excitations give rise to spin-triplet excitons. Due to the spin selection rule, spin-triplet excitons may be very stable in ferrimagnetic \ce{LiFe2F6}. Moreover, the spin-triplet excitons have spin to be 1 and -1 for the positive and negative ferroelectric polarization, respectively. More importantly, the spin-triplet excitons with S=1 and -1 can be switched by electric field due to strong magnetoelectric coupling. If the spin-triplet excitons can condense into superflow, the two spin superfluid phases with S=1 and -1 can be switched by electric field, which is very important for theory, experiment, and the design of new devices. Finaly, we also calculate the electronic structure of the ferrimagnetic \ce{LiFe2F6} by hybrid functional. The bandgap of the ferrimagnetic \ce{LiFe2F6} is increased to 1.08 $eV$, but the spins of the valence band and the conduction band are still opposite. Thus, the ferrimagnetic \ce{LiFe2F6} may still possess the spin-triplet excitonic phase in the framework of hybrid functional.  

In summary, based on symmetry analysis and the first-principles electronic calculations, we predict the \ce{LiFe2F6} is a $d$-wave altermagnetic, charge-ordering-mediated ferroelectric material. Under biaxial compressive strain, the \ce{LiFe2F6} transforms into charge-ordering-mediated ferrimagnetic, ferroelectric phase with strong magnetoelectric coupling. Interestingly, the spins of the valence band and the conduction band are opposite in ferrimagnetic \ce{LiFe2F6}, which facilitates the spin-triplet excitonic insulator phase. More importantly, If the spin-triplet excitons condense into superflow, the two spin superfluid phases with S=1 and -1 can be switched by electric field. Due to the abundance of novel physical properties, \ce{LiFe2F6} will certainly attract a wide range of theoretical and experimental interest.

\begin{acknowledgments}
We thank Z.-X. Liu, S. Qu and S. Dong for valuable discussions. This work was financially supported by the National Natural Science Foundation of China (No.11934020 and No.12204533). Computational resources have been provided by the Physical Laboratory of High Performance Computing at Renmin University of China.
\end{acknowledgments}

\nocite{*}

\bibliography{main}

\end{document}